\documentstyle[epsfig]{mn}
\begin{document}


\title{Do Type Ia Supernovae prove $\Lambda$ $>$ 0 ?}
\author
[Rowan-Robinson M.]{Michael Rowan-Robinson\\
$^1$Astrophysics Group, Imperial College London, Blackett Laboratory,
Prince Consort Road, London SW7 2BW}
\maketitle
\begin{abstract} 
The evidence for positive cosmological constant $\Lambda$ from Type Ia supernovae
is reexamined.

Both high redshift supernova teams are found to underestimate the effects of host galaxy extinction.
The evidence for an absolute magnitude- decay time relation is much weakened if supernovae not
observed before maximum light are excluded. Inclusion of such objects artificially supresses
the scatter about the mean relation.  

With a consistent treatment of host galaxy extinction and elimination of supernovae not
observed before maximum, the evidence for a positive lambda is not very significant (3-4 $\sigma$).  
A factor which may contribute to
apparent faintness of high z supernovae is evolution of the host galaxy extinction with z.

The Hubble diagram using all high z distance estimates, including SZ clusters and gravitational
lens time-delay estimates, does not appear inconsistent with an
$\Omega_o$ = 1 model.

Although a positive $\Lambda$ can provide an, albeit physically unmotivated, resolution of
the low curvature implied by CMB experiments and evidence that $\Omega_o <$ 1 from large-scale
structure, the direct evidence from Type Ia supernovae seems at present to be inconclusive.

\end{abstract}


\section{Introduction}

The claim that the measured brightnesses of Type Ia supernovae at redshifts 0.1 - 1.0 imply $\Lambda > 0$  
(Schmidt et al 1998, Garnavich et al 1998, Riess et al 1998, Perlmutter
et al 1999, Fillipenko and Riess 2000, Riess et al 2001, Turner and Riess 2001) has had a dramatic 
effect on cosmology.  The model with
$\lambda_o$ = 0.7, where $\lambda_o = \Lambda/3 H_o^2$, and $\Omega_0$ = 0.3 has
become a concensus model, believed to be consistent with most evidence from large-scale
stucture and CMB fluctuations.

In this paper I test the strength of the evidence that $\Lambda > 0$ and show that there
are inconsistencies in the way the supernovae data have been analyzed.  When these are
removed, the strength of the evidence for $\Lambda > 0$ is much diminished.

To set the scene, Fig 1 shows $B_{max}$ versus log (cz)  for 117 Type Ia supernovae since 1956
from the Barbon et al (1998) catalogue (excluding those labelled '*' which are discovery magnitudes only), 
together with published supernovae from the high z 
programmes, corrected for Galactic and internal extinction, but not for decay-time effects, together with 
predicted curves from an $\Omega_o$ = 1 model.  At first sight there is not an enormous
difference between the high z and low z supernovae, except that the latter seem to
show a larger scatter.  Fig 2 shows the same excluding less reliable data (flagged ':', in the Barbon 
et al catalogue, or objects with pg magnitudes only (Leibundgut et al (1991)), correcting for peculiar 
velocity effects (see section 3),
using the Phillips et al (1999) internal extinction correction (see section 2) 
where available, and deleting two objects for which the dust correction is $>$ 1.4 mag.  The scatter
for the low z supernovae appears to have been reduced.  Finally Fig 3 shows the
supernovae actually used by Perlmutter et al (1999).  Now the scatter in the low z
supernovae is not much different from the high z supernovae and a difference in absolute
magnitude between low z and high z supernovae, relative to an $\Omega_o$ = 1 model,
can be perceived.  However comparison with Fig 2
suggests that the low z supernovae used may be an abnormally luminous subset of all supernovae. 
We will return to this point in section 4.

Excellent recent reviews of Type Ia supernovae, which fully discuss whether they can be thought of as a
homogenous population, have been given by Branch (1998), Hildebrand and Niemeyer (2000) and Leibundgut 
(2000, 2001).  In this paper
I shall assume that they form a single population and that their absolute magnitude at
maximum light depends, at most, on a small number of parameters.  I do not, for example, consider the 
possibility of evolution, discussed by Drell et al (2000). 

Absolute magnitudes are quoted for $H_o$ = 100 throughout.

\begin{figure}
\epsfig{file=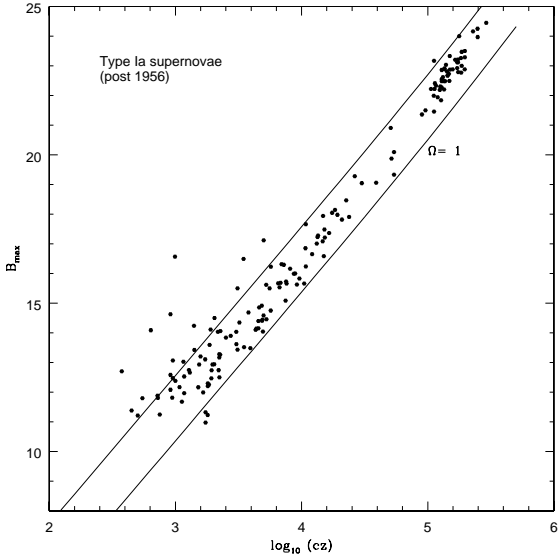,angle=0,width=8cm}
\caption{
$B_{max}$ versus log (cz) for all Type Ia supernovae post 1956.
Magnitudes are corrected for Galactic extinction, and for internal extinction (using the de 
Vaucouleurs prescription - see section 2).  A mean (B-V) colour of 0.0 has been assumed for
the supernovae for which only V at maximum light is known.  Solid curves are loci for $\Omega_o = 1$ model,
with $M_B = -19.64, -17.44 (H_o=100)$, corresponding to $<M_B> \pm 2 \sigma$ for solution (1) of Table 2.}
\end{figure}

\begin{figure}
\epsfig{file=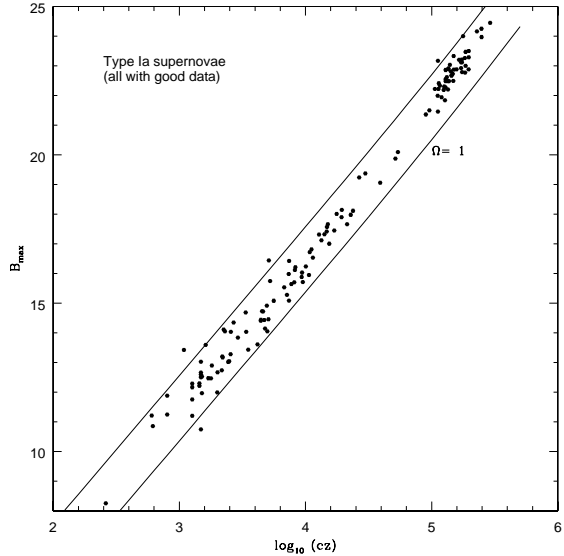,angle=0,width=8cm}
\caption{
$B_{max}$ versus log (cz) for Type Ia supernovae post 1956, excluding less reliable data,
correcting for effect of peculiar velocity, and
using the Phillips et al (1999) internal extinction correction where available.
Solid curves as for Fig 1.}
\end{figure}

\begin{figure}
\epsfig{file=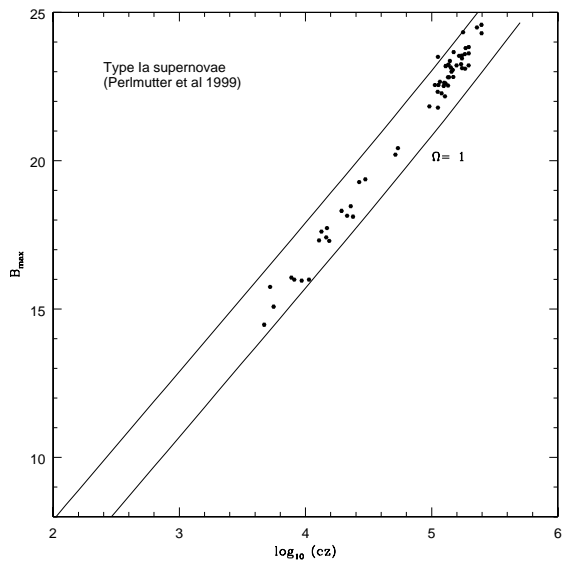,angle=0,width=8cm}
\caption{
$B_{max}$ versus log (cz) for Type Ia supernovae used by Perlmutter et al (1999).  No correction
for internal extinction is applied.
Solid curves are as for Fig 1, but shifted by +0.33 mag. to account for non-correction for extinction.}
\end{figure}

\section{Internal Extinction}

One of the most surprising claims of the high z supernovae teams is that internal
extinction in the high z supernovae is small or even negligible (Perlmutter et al 1999, Riess et al 2000).  
While some nearby Type Ia
supernovae take place in elliptical galaxies, where internal extinction in the host galaxy
may indeed be negligible, the majority take place in spiral galaxies, where internal
extinction can not be neglected.  Moreover as we look back towards z = 1, we know from
the CFRS survey that there is a marked increase in the fraction of star-forming
systems (Lilly et al 1996), and we would expect the average host galaxy extinction to be, if anything,
higher than in low z supernovae.

On average, the extinction along the line of sight to a (non edge-on) spiral galaxy
can be represented by de Vaucouleurs's prescription (de Vaucouleurs et al 1976):

\medskip

$A_{int}$ = 0.17 + $\alpha(T) lg_{10} (a/b)$, for T $>$ -4,  = 0 for T = -4, -5,

where T is the de Vaucouleurs galaxy type, a, b are the major and minor diameters of the galaxy,
and $\alpha$(T) = 0.2, 0.4, 0.6, 0.7, 0.8 for T = -3, -2, -1, 0, 1-8, respectively. 
I assume $lg_{10} (a/b)$ = 0.2 where this not known.


The extinction to a particularly supernova can be expected to show marked deviations from this
average value, since the dust distributions in galactic discs are very patchy.
Because of the cirrus-like distribution of interstellar dust clouds, lines-of-sight to some stars
will have much lower extinctions than this average value.  The presence of dense molecular clouds 
in the galaxy means that other lines-of-sight can have very much higher extinctions.  Phillips
et al (1999) have analyzed the extinction to 62 Type Ia supernovae, using both the colours at
maximum light and the colours at very late time, 30-90 days after maximum.  Their extinction corrections
do resolve a number of cases of anomalously faint Type Ia supernovae (eg 1995E, 1996Z, 1996ai). 
The agreement of their internal extinction values with those given by the de Vaucouleurs prescription 
is not brilliant in detail (see Table 1), but as expected give broadly the same median values 
(median $A_{int}$
for Phillips et al sample: 0.29, median de Vaucouleurs correction for same sample: 0.33).
These median values are broadly consistent with the Monte Carlo simulations of Hatano et al
(1998).
Figure 4 shows the correlation between $A_{int}$ and the (B-V) colour at maximum light, corrected for
Galactic extinction, with different symbols for Phillips et al (1999) estimates and those derived
from the de Vaucouleurs prescription.  The distribution is consistent with an intrinsic (unreddened) 
colour range of (B-V) = -0.1 to 0.1, combined with the usual  $A_{int}$ = 4.14 (B-V) relation. 

Riess et al (1998) have given estimates of the total extinction ($A_{int} + 4.14 E(B-V)_{Gal}$
for their sample of low supernova derived 
both via the MLCS method and via a set of templates (their Table 10).  
We can compare the template estimates directly with those of Phillips et al (1999) for the same 
galaxies (Fig 5).  The Riess
et al values are lower on average by 0.22 magnitudes, which implies that the set of templates 
(and the training set for the MLCS method) have not been completely dereddened.  The large scatter
in this diagram perhaps indicates the difficulty of estimating the host galaxy extinction for
supernovae.  If this
average underestimate of 0.22 mag. is added to the Riess et al estimates of  $A_{int}$ for 
high-z supernovae, the average value of
$A_{int}$ for these is almost identical to that for low z supernovae.
Thus the claim that high z supernovae have lower extinction than low z supernovae
seems to be based on a systematic underestimate of extinction in the high z galaxies.

Perlmutter et al (1999) estimate that host galaxy extinction is on average small both in local
and high z supernovae, and neglect it in most of their solutions.  
Parodi et al (2000) neglect internal extinction completely, 
preferring to cut out the redder objects (B-V$>$0.1) from their samples.  This will not change the 
relative absolute magnitudes between low and high z (or between supernovae with Cepheid
calibration and the others) provided the two samples end up
with the same mean extinction.  This, however, might be difficult to guarantee.  I have preferred
to correct the low z supernovae as described above, and then correct the Perlmutter et al data
by an average host galaxy extinction of 0.33 mag. 

Ellis and Sullivan (2001) have carried out HST imaging and Keck spectroscopy on host galaxies
for supernova used by Perlmutter et al (1999) and find that the Hubble diagram for supernovae in 
later type galaxies shows more scatter than those hosted by E$/$S0 galaxies, presumably due to
the effects of host galaxy extinction.



\begin{figure}
\epsfig{file=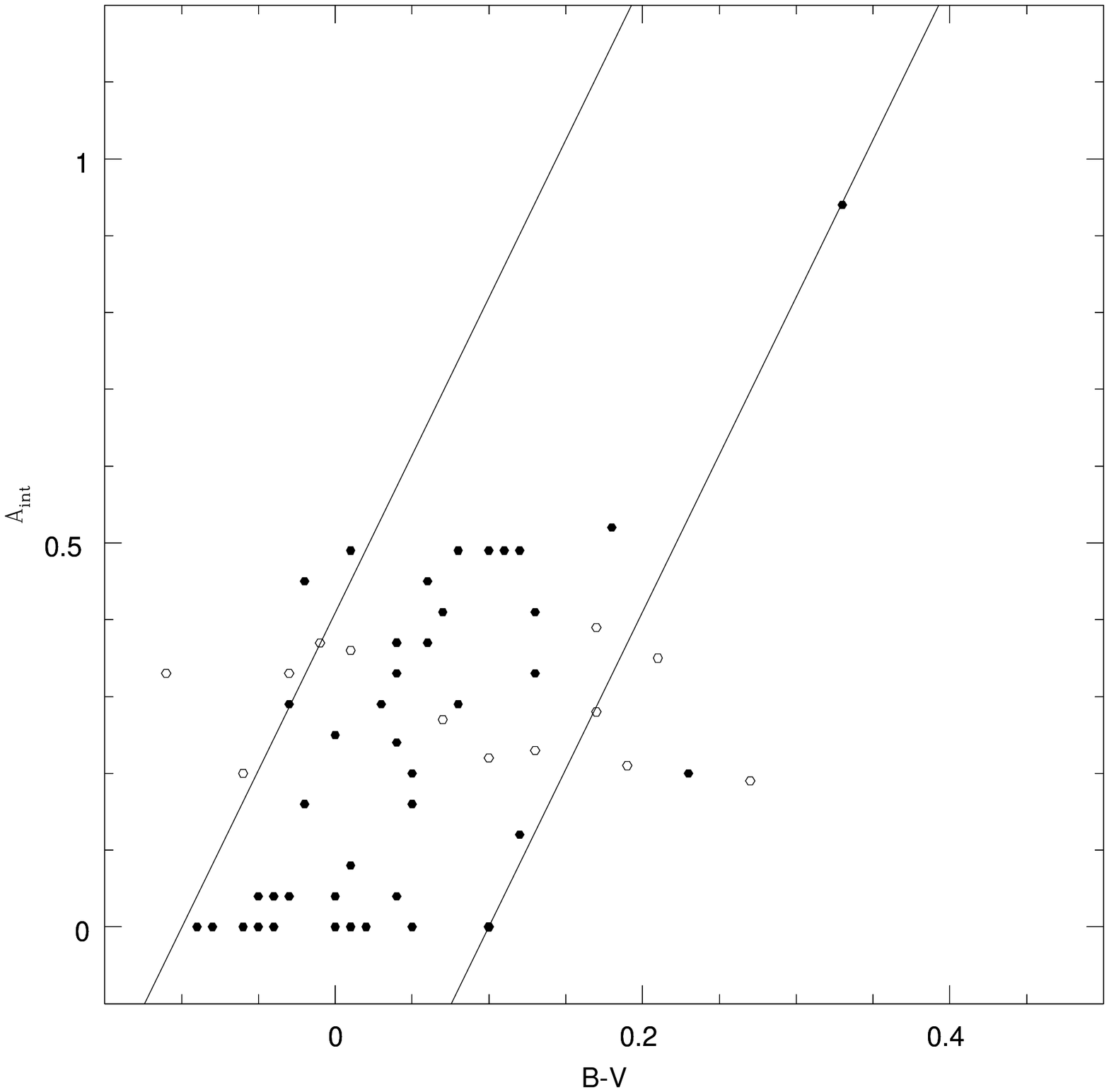,angle=0,width=8cm}
\caption{
$A_{int}$ versus $(B-V)$ (corrected for Galactic extinction) using Phillips et al (1999) extinction correction
filled circles) or de Vaucouleurs prescription (open circles).  The solid lines show the dust reddening
loci for intrinsic (unreddenened) colours of E(B-V) = -0.1 and 0.1.}
\end{figure}

\begin{figure}
\epsfig{file=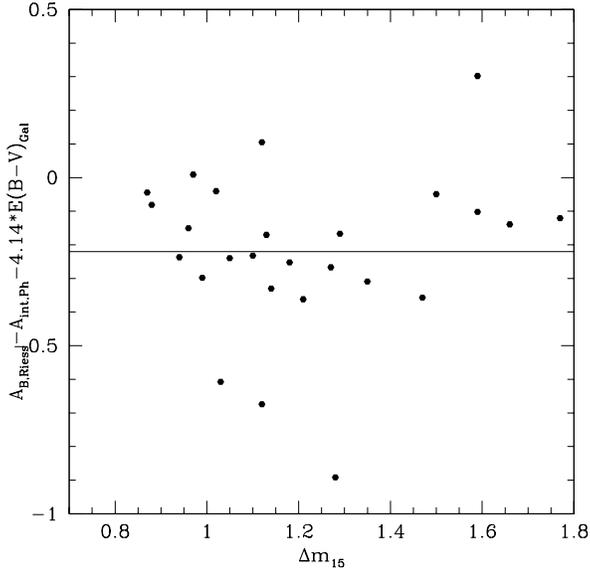, angle=0,width=8cm}
\caption{
Difference between the total extinction estimated by Riess et al (1998) and the sum of
the host galaxy extinction estimated by Phillips et al (1999) and the Galactic extinction,
plotted against $\Delta m_{15}$.}
\end{figure}


\section{Correction for peculiar velocity effects}
For nearby supernovae it is important to correct for the peculiar velocity of the host galaxy.
Hamuy et al (1995) tackle this be setting a minimum recession velocity of 2500 $km/s$ for their
local sample.  However since peculiar velocities can easily be in excess of 500 $km/s$ this
is probably not sufficient for an accurate result.  Here I have used the model for the
local (z $<$ 0.1) velocity field developed to analyze the CMB dipole, using data from the
IRAS PSCz redshift survey (Rowan-Robinson et al 2000).  I estimate the velocity error in
estimates from this model to be  100 + 0.2 x V  $km/s$ for V $<$ 15000 $km/s$, 400 $km/s$ for
V $>$ 15000 $km/s$.  These errors are incorporated into the subsequent analysis (points are weighted by
error$^{-2}$).
The code for this peculiar velocity model
is available at http://astro.ic.ac.uk/$\sim$mrr.

Tables 1 and 2 gives data for the Type Ia supernovae used in the present study.  The columns are as follows:
(1) supernova name, (2) host galaxy recession velocity, (3) same, corrected for peculiar velocity,
(4) blue magnitude at maximum light $B_{max}$,
(5) $E(B-V)_{Gal}$, from Schlegel et al (1998), (6) $A_{int}$ from de Vaucouleurs et al (1976) presciption,
(7) $A_{int}$ from Phillips et al (1999), (8) absolute magnitude $M_B$ ($\Omega_o$ = 1 universe), 
(9) distance modulus, assuming $M_B$ = -19.47 (Gibson et al 2000), (10) $\Delta m_{15}$, (11) $(B-V)_o$.
Sources of $B_{max}$, in order of preference: Hamuy et al (1996)/Perlmutter et al (1999)/Riess et al (1998,
Barbon et al (1999).  Sources of $\Delta m_{15}$, in order of preference: Phillips et al (1999),
Saha et al (1999), Riess et al (1998), Parodi et al (2001).  Sources of $(B-V)_o$: Hamuy et al (1996), 
Saha et al (1999), Parodi et al (2001), Leibundgut et al (1991).

\section{Absolute magnitude-decay time relation}

Once we have corrected for the effects of internal extinction, we can test whether the absolute
magnitude of supernovae at maximum light depends on decay time, or some other parameter like
the colour at maximum light $(B-V)_o$, as proposed by Tripp and Branch (1999), and Parodi et al (2000).

Phillips (1993) proposed that there is a strong dependence of absolute magnitude at maximum
light and the decay-time, characterized by the blue magnitude change during the 15 days after
maximum, $\Delta m_{15}$.  Specifically he found $d M_B/d \Delta m_{15}$ = 2.70.  Riess et al (1995) 
showed that the dispersion in the sn Ia Hubble 
diagram was significantly reduced by applying this correction.  Tammann and Sandage (1995)
used supernovae in galaxies for which distance estimates were available to set a strong limit
on the slope of the $M_B - \Delta m_{15}$ relation ($<$ 0.88).  Hamuy et al (1996) used a new sample
of well-studied supernovae to derive a B-band slope of 0.784 $\pm$ 0.18 for the
$M_B - \Delta m_{15}$ relation, consistent with the Tammann and Sandage limit, and much lower 
than the original claim of Phillips (1993).  Riess et al (1996) have discussed a related
method of analyzing this correlation, the MLCS method.

However there is a further consideration here.  It is really only valid to carry out this analysis 
on supernovae which have been detected prior to maximum light.  The process used hitherto 
by all workers in this field of extrapolating to maximum light assuming an $M_B - \Delta m_{15}$,
or equivalently MLCS, relation is assuming that all extrapolated objects adhere to the mean line 
of the relation, thus underestimating the scatter in the relation.  This is a process which
artificially improves the apparent signal-to-noise of the final Hubble relation or
$\Lambda > 0$ signal.  Hamuy et al (1995) do make some allowance for this in assigning
a larger uncertainty (and hence lower weight) to supernova first observed after maximum.

Fig 6 shows a plot of the absolute magnitude at maximum light, $M_B$, corrected by
0.784  $\Delta m_{15}$, versus $(B-V)_o$, the colour at maximum light corrected for
the effects of extinction, using the Phillips et al (1999) estimates of extinction.
No clear correlation remains between these corrected quantities.  Thus correction
for the $M_B - \Delta m_{15}$ relation removes most of the correlation between
$M_B$ and $(B-V)_o$.

Fig 7 shows the relations between $M_B$, corrected for extinction, and $\Delta M_{15}$,
for objects detected at least 1 day prior to maximum light.  The best fit linear relation is shown,
which has slope 0.99 $\pm$0.38, consistent with values reported by Hamuy et al (1995), 0.85
$\pm$0.13, and by Hamuy et al (1996), 0.78 $\pm$ 0.18.  However the significance
of the relation is reduced, because of the smaller number of data points, and is now only
2.6 $\sigma$.  The rms deviation from this mean relation is 0.44 mag, much larger than is generally
claimed for this relation.  For example, Riess et al (1996) claim that the residual sigma
after correction for the $M_B - \Delta m_{15}$ relation is 0.12 mag.   

The spurious reduction in scatter generated by extrapolating supernovae first observed
after maximum can be seen in Fig 8, which shows the same relation for supernovae first
observed after maximum.  The same effect can in fact be seen in the top panel of Fig 4
in Parodi et al (2000). The marked difference between Fig 7 and 8 suggests that 
supernovae first observed after maximum should be excluded from the analysis.

Perlmutter et al (1999) have applied a different version of the $M_B - \Delta m_{15}$ method, which
they call the 'stretch' method, to high z supernovae.  This method appears to give a significantly 
lower correction to $M_B$ as a function of $\Delta m_{15}$ (Leibundgut 2000). Figure 9 shows the Perlmutter et al
'stretch' correction to the absolute magnitude, as a function of $\Delta m_{15}$, for 18 low z
supernovae (note that they omit the two supernovae with $\Delta m_{15}$ = 1.69 from their solution).  
The slope is 0.275 $\pm$0.04, only one third of the Hamuy et al (1996) value.  The method has been
further discussed by Efstathiou et al (1999) and Goldhaber et al (2001), but no explanation
for the lower slope is given.

\begin{figure}
\epsfig{file=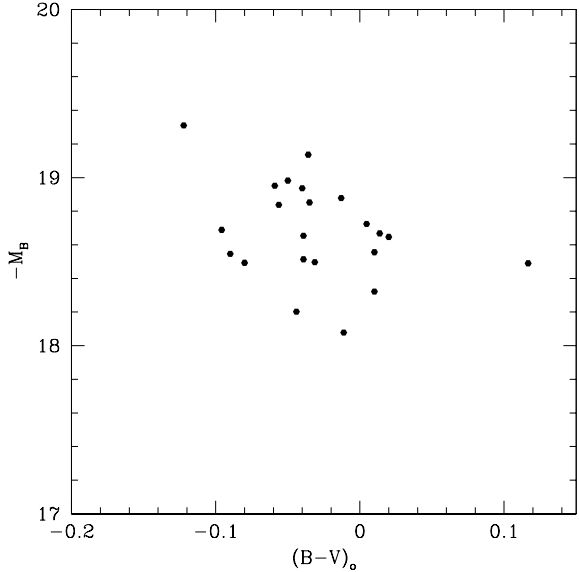,angle=0,width=8cm}
\caption{
$M_B$ at maximum light, corrected for extinction (using Phillips et al prescription) and by
0.784  $\Delta m_{15}$ , versus $(B-V)_o$.}
\end{figure}


\begin{figure}
\epsfig{file=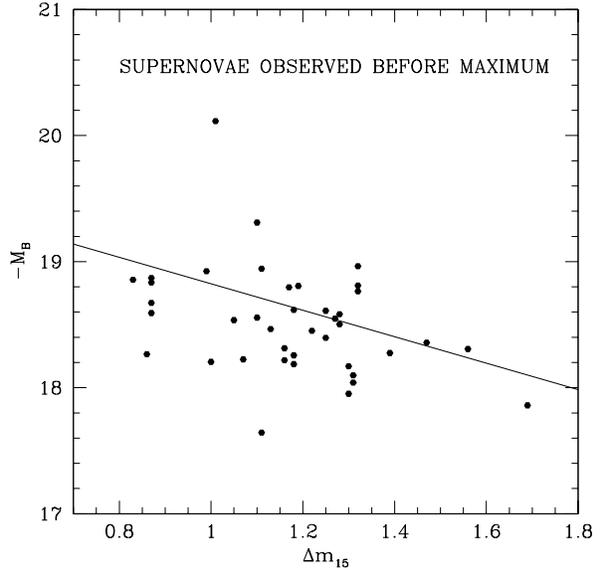,angle=0,width=8cm}
\caption{$M_B$ at maximum light, corrected for extinction, versus $\Delta m_{15}$,
for supernovae first observed before maximum.  The best-fitting line has slope -0.99,
but the data are also consistent with no correlation.}
\end{figure}

\begin{figure}
\epsfig{file=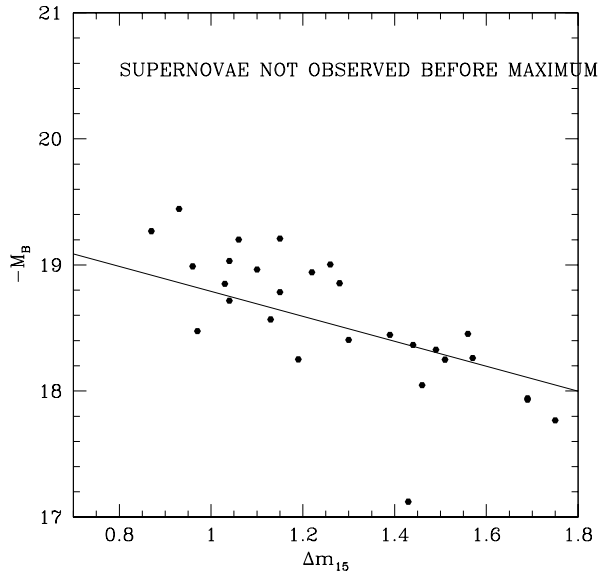,angle=0,width=8cm}
\caption{$M_B$ at maximum light, corrected for extinction, versus $\Delta m_{15}$,
for supernovae not observed until after maximum.  Note the (spurious)
reduction in scatter.}
\end{figure}

\begin{figure}
\epsfig{file=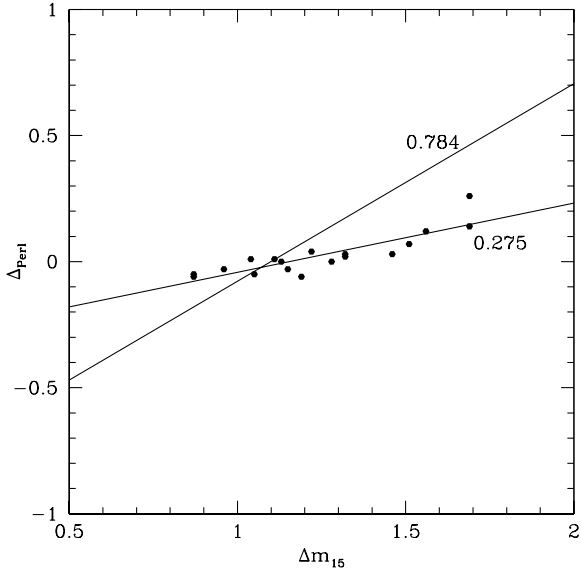,angle=0,width=8cm}
\caption{
Perlmutter et al 'stretch' correction to $M_B$, $\Delta_{Perl}$, versus $\Delta m_{15}$,
for low z supernovae.  The best fit slope is 0.275, well below the value of 0.784 quoted by
Hamuy et al (1996).}
\end{figure}

\section{Testing for positive $\Lambda$}

With the criteria established above (i) that a full and consistent correction must be made for extinction, 
(ii) that if the the $M_B - \Delta m_{15}$ relation is used it should only be applied to supernovae detected
before maximum light, we can now reexamine the Hubble diagram for supernovae.  I consider several different 
samples:

(1) all well-observed supernovae, with no correction for the $M_B - \Delta m_{15}$ relation.  Supernovae with
only pg magnitudes are excluded, as also are supernovae first observed after maximum.  The de
Vaucouleurs extinction correction is used for supernovae not studied by Phillips et al (1999).  Supernovae
with $A_{int} >$ 1.4 are excluded.

(2) all well-observed supernovae for which in addition  $\Delta m_{15}$ is known, with correction for 
the $M_B - \Delta m_{15}$ relation, using the 0.784 slope of Hamuy et al (1966).

(3) as (2), but with internal extinction set to zero, as advocated by Perlmutter et al (1999).

(4) as (2), but using the Perlmutter 'stretch' correction (or 0.275 $\Delta m_{15}$ where stretch 
correction not available).

(5) as (2) but using the quadratic $\Delta m_{15}$ correction of Phillips et al (1999).

It was not possible to independently check the effect of applying the MLCS method for correcting for
decay-time correlations because the set of training vectors published by Riess et al (1996) is not
the one actually being used in the high z supernova analysis (Riess et al 1998).

The mean absolute magnitudes for low z ( z $<$ 0.1) and high z supernovae, in an Einstein de Sitter model
are tabulated for the different samples in Table 3.  Without the correction for $\Delta m_{15}$, the 
significance of the difference in absolute magnitude for 53 low and 52 high redshift supernovae is 
only 2.8 $\sigma$, hardly sufficient to justify the current wide acceptance of positive $\Lambda$. 
The significance increases to 3.5 $\sigma$ if only the 26 supernovae observed since 1990 are used. 

Including the $\Delta m_{15}$ correction, the significance increases to 4.0 $\sigma$ (3.9 $\sigma$ if we
use the Perlmutter correction, 4.6 $\sigma$ if we use the quadratic $\Delta m_{15}$ 
correction of Phillips et al 1999, which is the form used by Riess et al 1998).  
Could this increase in significance be due to some bias in the local supernova sample for which
$\Delta m_{15}$ has been measured ?  The mean absolute magnitude at maximum light, corrected for extinction 
but not for $\Delta m_{15}$, for all 53 'good' low z supernovae is -18.54.  The mean for those for 
which $\Delta m_{15}$ has been
measured is -18.64, 0.1 magnitudes brighter.  In fact the difference in the mean between those for which
$\Delta m_{15}$ has been measured and those for which it has not been measured is 0.23 mag., almost the 
size of the whole signal on which the claim for positive $\Lambda$ is based.  However there is no
difference in mean absolute magnitude between the mean absolute magnitude for the 10 local Calan Tololo
supernovae used by by the high-z supernova teams and the 16 other local supernovae studied since 1990,
so there is no evidence that the Calan Tololo sample is biased.  An alternative explanation of the
fainter mean absolute magnitudes seen in the pre-1990 data is a systematic error in the photographic
photometry used.

There is another factor which may contribute to the apparent faintness of high z supernovae.  Current models of the
star formation history of the universe, which show a strong peak in the star formation rate at z = 1-2,
imply that the mean dust optical depth in galaxies would be expected to increase with redshift out to
z =1.5-2 (Calzetti and Heckman 1999, Pei et al 1999).  Line 6 of Table 3 shows the effect of adopting
Calzetti and Heckman's model 3 for the evolution of $A_{int}(z)$, a model which agrees well with models for
infrared and submillimetre sourcounts and backgrounds (eg Rowan-Robinson 2001a) and direct evidence
from oberved estimates of the star formation history (Rowan-Robinson 2001b).  This correction, which increases
the average $A_{int}$ for the high z supernovae by 0.15 mag., is sufficient to reduce the significance
of the magnitude difference between high and low z supernovae, in an Einstein de Sitter model, to 3.0 and 1.5
$\sigma$ for cases with and without correction for $\Delta m_{15}$.

\section{Hubble diagram at high redshift}
Rowan-Robinson (2001c) has reviewed distance estimates and the Hubble constant.  In addition
to Type Ia supernovae, S-Z clusters and gravitational lens time-delay methods also give distance
estimates at z $>$ 0.1.  Figure 10 shows a compilation of all these high z estimates.  I have also
included the HDF supernova (1997ff) at the redshift 1.7, proposed by Riess et al (2001)
although Rowan-Robinson (2001b) finds a photometric redshift z = 1.4 for the parent galaxy.  The
assumed B band magnitude at maximum light is 25.5 (Fig 7 of Riess et al)  and the mean absolute 
magnitude for Type Ia
supernovae is assumed to be  -19.47 (Gibson et al 2000).  The model with $\Omega_o$ = 1
is a good overall fit to these data and the HDF supernova lies right on the $\Omega_o$ = 1 mean line.
A least squares fit to these data, with an assumed $H_o$ of 63 $km/s/Mpc$ (Rowan-Robinson 2001c) and
the spatial curvature parameter k = 0,
yields  $\Omega_o = 0.81 \pm 0.12$, where the error in the distance modulus has been taken to be
0.35 mag. for all points. 

\begin{table*}
\caption{Data for low redshift Type Ia supernovae, excluding supernovae not observed before maximum}
\begin{tabular}{llllllllllll}
name & $V$ & $V_{corr}$ & $B_{max}$ & $E(B-V)_{Gal}$ & $A_{int,V}$ & $A_{int,Ph}$ &
$M_B$ & $\mu_o$ & $\Delta m_{15}$ & $(B-V)_o$ \\
\bf{low z}& & & & & & & & & & & \\
1960R   &   751 &   802 & 11.60 & 0.0255 & 0.25 & -   & -18.28 &  30.71 & - &  0.17\\
1963P   &  1414 &  1088 & 14.00 & 0.0445 & 0.39 & -   & -16.76 &  32.90 & - &  -  \\
1965I   &  1172 &  1511 & 12.41 & 0.0198 & 0.36 & -   & -18.93 &  31.44 & - & -0.17\\
1970J   &  3791 &  3360 & 15.00 & 0.0760 & 0.00 & -   & -17.95 &  34.16 & 1.30 &  0.10\\
1971I   &   503 &   604 & 11.60 & 0.0120 & 0.34 & -   & -17.70 &  30.68 & - &  0.17\\
1971L   &  1659 &  1996 & 13.00 & 0.1505 & 0.39 & -   & -19.52 &  31.46 & - &  0.17\\
1972E   &   403 &   261 &  8.49 & 0.0483 & 0.42 & 0.04 & -18.83 &  27.34 & 0.87 & -0.03\\
1972J   &  3213 &  2690 & 14.76 & 0.0468 & 0.22 & -   & -17.81 &  33.82 & - &  0.10\\
1974G   &   720 &   802 & 12.28 & 0.0124 & 0.35 & -   & -17.64 &  31.35 & 1.11 &  0.21\\
1974J   &  7468 &  7390 & 15.60 & 0.0692 & 0.23 & -   & -19.27 &  34.55 & - &  -0.28\\
1975G   &  1900 &  2246 & 14.44 & 0.0092 & 0.29 & -   & -17.65 &  33.58 & - &  -  \\
1975N   &  1867 &  1620 & 14.00 & 0.0302 & 0.28 & -   & -17.46 &  33.06 & - &  0.17\\
1976J   &  4558 &  4164 & 14.28 & 0.0263 & 0.56 & -   & -19.49 &  33.08 & 0.90 &  0.02\\
1978E   &  4856 &  4423 & 15.40 & 0.1612 & 0.33 & -   & -18.83 &  33.87 & - &  -0.19\\
1979B   &   954 &  1699 & 12.70 & 0.0099 & 0.19 & -   & -18.68 &  31.94 & - &  0.27\\
1980N   &  1806 &  1442 & 12.50 & 0.0207 & 0.21 & 0.20 & -18.58 &  31.67 & 1.28 &  0.05\\
1981B   &  1804 &  1266 & 11.74 & 0.0207 & 0.43 & 0.45 & -19.31 &  30.69 & 1.10 & -0.02\\
1981D   &  1806 &  1442 & 12.59 & 0.0207 & 0.21 & -   & -18.50 &  31.76 & 1.28 &  0.19\\
1983G   &  1172 &  1511 & 12.97 & 0.0198 & 0.36 & -   & -18.37 &  32.00 & - &  0.01\\
1983U   &  1156 &  1483 & 13.40 & 0.0216 & 0.29 & -   & -17.84 &  32.49 & - &  -  \\
1983W   &  1937 &  2171 & 13.30 & 0.0104 & 0.52 & -   & -18.95 &  32.21 & - &  -  \\
1986A   &  3039 &  3385 & 14.40 & 0.0409 & 0.20 & -   & -18.62 &  33.50 & - &  -  \\
1987D   &  2227 &  2551 & 13.70 & 0.0217 & 0.33 & -   & -18.76 &  32.75 & - &  -  \\
1988F   &  5274 &  5090 & 14.80 & 0.0223 & 0.25 & -   & -19.08 &  33.93 & - &  -  \\
1989A   &  2514 &  2914 & 14.10 & 0.0221 & 0.17 & -   & -18.49 &  33.31 & - &  -  \\
1989B   &   726 &   617 & 12.34 & 0.0304 & 0.41 & 1.36 & -18.10 &  31.27 & 1.31 &  -  \\
1989M   &  1518 &  1266 & 12.56 & 0.0376 & 0.24 & -   & -18.35 &  31.63 & - &  -  \\
1990N   &   998 &  1266 & 12.76 & 0.0243 & 0.28 & 0.37 & -18.23 &  31.85 & 1.07 &  0.04\\
1990Y   & 10800 & 10804 & 17.70 & 0.0095 & 0.00 & 0.94 & -18.47 &  37.13 & 1.13 &  0.33\\
1990af  & 15180 & 14807 & 17.87 & 0.0336 & 0.25 & 0.16 & -18.31 &  36.95 & 1.56 &  0.05\\
1991M   &  2169 &  2562 & 14.55 & 0.0461 & 0.33 & -   & -17.89 &  33.63 & - &  -  \\
1992A   &  1845 &  1484 & 12.56 & 0.0138 & 0.25 & 0.00 & -18.36 &  31.72 & 1.47 &  0.02\\
1992G   &  1580 &  2193 & 13.63 & 0.0121 & 0.38 & -   & -18.41 &  32.77 & - &  -  \\
1992P   &  7616 &  8159 & 16.08 & 0.0210 & 0.33 & 0.29 & -18.87 &  35.13 & 0.87 & -0.03\\
1992ag  &  7544 &  7724 & 16.41 & 0.0856 & 0.33 & 0.41 & -18.81 &  35.20 & 1.19 &  0.08\\
1992al  &  4377 &  4711 & 14.60 & 0.0315 & 0.33 & 0.04 & -18.94 &  33.61 & 1.11 & -0.05\\
1992bc  &  5700 &  5602 & 15.16 & 0.0196 & 0.33 & 0.00 & -18.67 &  34.22 & 0.87 & -0.08\\
1992bh  & 13500 & 13384 & 17.70 & 0.0217 & 0.33 & 0.49 & -18.54 &  36.75 & 1.05 &  0.08\\
1992bo  &  5575 &  5245 & 15.86 & 0.0270 & 0.25 & 0.00 & -17.86 &  34.97 & 1.69 &  0.01\\
1992bp  & 23790 & 23791 & 18.41 & 0.0717 & 0.21 & 0.00 & -18.81 &  37.37 & 1.32 & -0.05\\
1993L   &  1925 &  1784 & 13.20 & 0.0179 & 0.66 & -   & -18.62 &  32.11 & 1.18 &  -  \\
1993O   & 15300 & 14750 & 17.67 & 0.0610 & 0.21 & 0.00 & -18.45 &  36.68 & 1.22 & -0.09\\
1993ag  & 14700 & 15417 & 17.72 & 0.1031 & 0.21 & 0.29 & -18.96 &  36.55 & 1.32 &  0.03\\
1994D   &   450 &  1266 & 11.84 & 0.0217 & 0.37 & 0.00 & -18.76 &  30.85 & 1.32 & -0.04\\
1994S   &  4550 &  4523 & 14.80 & 0.0170 & 0.33 & 0.00 & -18.56 &  33.87 & 1.10 &  0.01\\
1994ae  &  1282 &  1483 & 13.20 & 0.0284 & 0.34 & 0.49 & -18.27 &  32.21 & 0.86 &  0.10\\
1995D   &  1967 &  2442 & 13.40 & 0.0533 & 0.25 & 0.16 & -18.92 &  32.40 & 0.99 & -0.02\\
1995al  &  1541 &  2011 & 13.36 & 0.0207 & 0.34 & 0.61 & -18.86 &  32.40 & 0.83 &  -  \\
1996X   &  2032 &  1811 & 13.22 & 0.0685 & 0.00 & 0.04 & -18.40 &  32.41 & 1.25 & -0.03\\
1996bl  & 10800 & 10063 & 17.05 & 0.1173 & 0.33 & 0.33 & -18.80 &  35.70 & 1.17 &  0.04\\
1996bo  &  5241 &  4617 & 16.17 & 0.0724 & 0.25 & 1.15 & -18.61 &  35.09 & 1.25 &  -  \\
1997ej  &  6686 &  6774 & 15.85 & 0.0170 & 0.25 & -   & -18.64 &  35.00 & - &  -  \\
1998bu  &   943 &  1483 & 12.22 & 0.0303 & 0.28 & 1.35 & -20.11 &  31.28 & 1.01 &  -  \\
\end{tabular}
\end{table*}

\begin{table*}
\caption{Data for high redshift Type Ia supernovae}
\begin{tabular}{llllllllllll}
name & $V$ & $V_{corr}$ & $B_{max}$ & $E(B-V)_{Gal}$ & $A_{int,dV}$ & $A_{int,Ph}$ &
$M_B$ & $\mu_o$ & $\Delta m_{15}$ & $(B-V)_o$ \\
& & & & & & & & & & & \\
\bf{Riess 98}& & & & & & & & & & & \\
1996E   &     0 &128914 & 22.81 & 0.0000 & - & 0.10 & -18.26 &  41.95 & 1.18 &  -  \\
1996H   &     0 &185876 & 23.23 & 0.0000 & - & 0.00 & -18.59 &  42.37 & 0.87 &  -  \\
1996I   &     0 &170886 & 23.35 & 0.0000 & - & 0.00 & -18.27 &  42.49 & 1.39 &  -  \\
1996J   &     0 & 89940 & 22.23 & 0.0000 & - & 0.64 & -18.55 &  41.37 & 1.27 &  -  \\
1996K   &     0 &113924 & 22.64 & 0.0000 & - & 0.00 & -18.04 &  41.78 & 1.31 &  -  \\
1996U   &     0 &128914 & 22.78 & 0.0000 & - & 0.00 & -18.19 &  41.92 & 1.18 &  -  \\
1997ce  &     0 &131912 & 22.85 & 0.0000 & - & 0.00 & -18.17 &  41.99 & 1.30 &  -  \\
1997cj  &     0 &149900 & 23.19 & 0.0000 & - & 0.09 & -18.22 &  42.33 & 1.16 &  -  \\
1997ck  &     0 &290806 & 24.78 & 0.0000 & - & 0.10 & -18.20 &  43.92 & 1.00 &  -  \\
1995K   &     0 &143904 & 22.91 & 0.0000 & - & 0.00 & -18.31 &  42.05 & 1.16 &  -  \\
\bf{Perlmutter 99}& & & & & & & & & & & \\
1992bi  &     0 &137308 & 22.81 & 0.0000 & 0.33 & -   & -18.40 &  41.95 & - &  -  \\
1994F   &     0 &106129 & 22.55 & 0.0000 & 0.33 & -   & -18.07 &  41.69 & - &  -  \\
1994G   &     0 &127415 & 22.17 & 0.0000 & 0.33 & -   & -18.87 &  41.31 & - &  -  \\
1994H   &     0 &112125 & 21.79 & 0.0000 & 0.33 & -   & -18.95 &  40.93 & - &  -  \\
1994al  &     0 &125916 & 22.63 & 0.0000 & 0.33 & -   & -18.38 &  41.77 & - &  -  \\
1994am  &     0 &111526 & 22.32 & 0.0000 & 0.33 & -   & -18.41 &  41.46 & - &  -  \\
1994an  &     0 &113324 & 22.55 & 0.0000 & 0.33 & -   & -18.22 &  41.69 & - &  -  \\
1995aq  &     0 &135809 & 23.24 & 0.0000 & 0.33 & -   & -17.95 &  42.38 & - &  -  \\
1995ar  &     0 &139407 & 23.36 & 0.0000 & 0.33 & -   & -17.89 &  42.50 & - &  -  \\
1995as  &     0 &149300 & 23.66 & 0.0000 & 0.33 & -   & -17.75 &  42.80 & - &  -  \\
1995at  &     0 &196369 & 23.21 & 0.0000 & 0.33 & -   & -18.84 &  42.35 & - &  -  \\
1995aw  &     0 &119920 & 22.27 & 0.0000 & 0.33 & -   & -18.63 &  41.41 & - &  -  \\
1995ax  &     0 &184377 & 23.10 & 0.0000 & 0.33 & -   & -18.80 &  42.24 & - &  -  \\
1995ay  &     0 &143904 & 23.00 & 0.0000 & 0.33 & -   & -18.32 &  42.14 & - &  -  \\
1995az  &     0 &134910 & 22.53 & 0.0000 & 0.33 & -   & -18.64 &  41.67 & - &  -  \\
1995ba  &     0 &116322 & 22.66 & 0.0000 & 0.33 & -   & -18.17 &  41.80 & - &  -  \\
1996cf  &     0 &170886 & 23.25 & 0.0000 & 0.33 & -   & -18.48 &  42.39 & - &  -  \\
1996cg  &     0 &146902 & 23.06 & 0.0000 & 0.33 & -   & -18.31 &  42.20 & - &  -  \\
1996ci  &     0 &148401 & 22.82 & 0.0000 & 0.33 & -   & -18.58 &  41.96 & - &  -  \\
1996ck  &     0 &196669 & 23.62 & 0.0000 & 0.33 & -   & -18.43 &  42.76 & - &  -  \\
1996cl  &     0 &248234 & 24.58 & 0.0000 & 0.33 & -   & -18.03 &  43.72 & - &  -  \\
1996cm  &     0 &134910 & 23.22 & 0.0000 & 0.33 & -   & -17.95 &  42.36 & - &  -  \\
1996cn  &     0 &128914 & 23.19 & 0.0000 & 0.33 & -   & -17.88 &  42.33 & - &  -  \\
1997F   &     0 &173884 & 23.45 & 0.0000 & 0.33 & -   & -18.32 &  42.59 & - &  -  \\
1997G   &     0 &228747 & 24.49 & 0.0000 & 0.33 & -   & -17.92 &  43.63 & - &  -  \\
1997H   &     0 &157695 & 23.21 & 0.0000 & 0.33 & -   & -18.33 &  42.35 & - &  -  \\
1997I   &     0 & 51566 & 20.20 & 0.0000 & 0.33 & -   & -18.78 &  39.34 & - &  -  \\
1997J   &     0 &185576 & 23.80 & 0.0000 & 0.33 & -   & -18.12 &  42.94 & - &  -  \\
1997K   &     0 &177482 & 24.33 & 0.0000 & 0.33 & -   & -17.48 &  43.47 & - &  -  \\
1997L   &     0 &164890 & 23.53 & 0.0000 & 0.33 & -   & -18.11 &  42.67 & - &  -  \\
1997N   &     0 & 53964 & 20.42 & 0.0000 & 0.33 & -   & -18.66 &  39.56 & - &  -  \\
1997O   &     0 &112125 & 23.50 & 0.0000 & 0.33 & -   & -17.24 &  42.64 & - &  -  \\
1997P   &     0 &141506 & 23.14 & 0.0000 & 0.33 & -   & -18.14 &  42.28 & - &  -  \\
1997Q   &     0 &128914 & 22.60 & 0.0000 & 0.33 & -   & -18.47 &  41.74 & - &  -  \\
1997R   &     0 &196969 & 23.83 & 0.0000 & 0.33 & -   & -18.23 &  42.97 & - &  -  \\
1997S   &     0 &183478 & 23.59 & 0.0000 & 0.33 & -   & -18.30 &  42.73 & - &  -  \\
1997ac  &     0 & 95936 & 21.83 & 0.0000 & 0.33 & -   & -18.56 &  40.97 & - &  -  \\
1997af  &     0 &173584 & 23.54 & 0.0000 & 0.33 & -   & -18.22 &  42.68 & - &  -  \\
1997ai  &     0 &134910 & 22.81 & 0.0000 & 0.33 & -   & -18.36 &  41.95 & - &  -  \\
1997aj  &     0 &174184 & 23.12 & 0.0000 & 0.33 & -   & -18.65 &  42.26 & - &  -  \\
1997am  &     0 &124717 & 22.52 & 0.0000 & 0.33 & -   & -18.47 &  41.66 & - &  -  \\
1997ap  &     0 &248834 & 24.30 & 0.0000 & 0.33 & -   & -18.31 &  43.44 & - &  -  \\
\end{tabular}
\end{table*}

\begin{table*}
\caption{Mean difference in absolute magnitude for low and high z supernovae, in ES model}
\begin{tabular}{llllllllllll}
sample & no.  & no. & $<A_{int}>$ & $<A_{int}>$ & $<M_B>$ & $\sigma$ & $<M_B>$ & $\sigma$ &
$\Delta M_B$ & $\sigma_{diff}$ & $\Delta/\sigma_{diff}$\\
& low z & high z & low z & high z & low z & & high z & & & & \\
& & & & & & & & & \\
(1) all good sn,  & 53 & 52 & 0.327 & 0.329 & -18.54 & 0.55 & -18.30 & 0.33 & 0.25 & 0.089 & 2.8 \\
no $\Delta M_{15}$ corrn & & & & & & & & & & & \\
& & & & & & & & & & & \\
(2) with $\Delta M_{15}$ corrn & 31 & 52 & 0.353 & 0.329 & -18.69 & 0.42 & -18.30 & 0.38 & 0.39 & 0.093 & 4.25 \\
& & & & & & & & & & & \\
(3) with $\Delta M_{15},$ corrn & 31 & 52 & 0.0 & 0.0 & -18.35 & 0.48 & -17.975 & 0.38 & 0.375 & 0.101 & 3.7 \\
but no dust correctn & & & & & & & & & & & \\
& & & & & & & & & & & \\
(4) with Perlmutter corrn & 31 & 52 & 0.353 & 0.329 & -18.655 & 0.44 & -18.30 & 0.33 & 0.36 & 0.092 & 3.9 \\
& & & & & & & & & & & \\
(5) with quadratic  & 31 & 52 & 0.353 & 0.329 & -18.72 & 0.41 & -18.31 & 0.38 & 0.41 & 0.091 & 4.6 \\
$\Delta M_{15}$ corrn & & & & & & & & & & & \\
& & & & & & & & & & & \\
(6) with $\Delta M_{15}$ corrn & 30 & 52 & 0.283 & 0.446 & -18.69 & 0.42 & -18.42 & 0.37 & 0.27 & 0.091 & 3.0 \\
and dust evoln & & & & & & & & & & & \\ 
(7) with no $\Delta M_{15}$ corrn & 53 & 52 & 0.331 & 0.446 & -18.54 & 0.55 & -18.41 & 0.33 & 0.13 & 0.089 & 1.5 \\
and dust evoln & & & & & & & & & & & \\
\end{tabular}
\end{table*}

\begin{figure*}
\epsfig{file=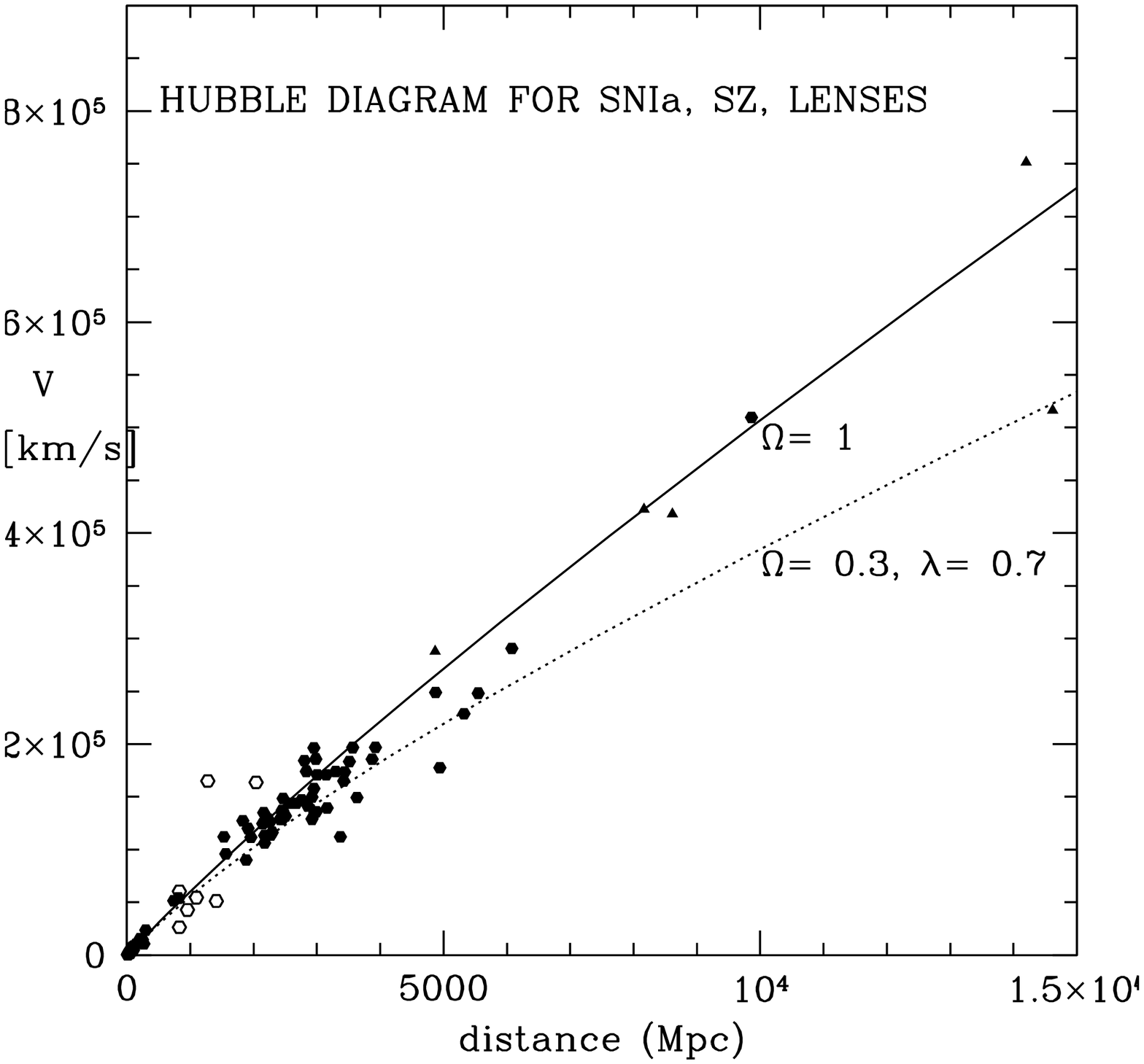,angle=0,width=13cm}
\caption{
Hubble diagram for Type Ia supernovae (filled circles), gravitational lenses (filled triangles) and SZ clusters
(open circles).}
\end{figure*}

\section{Discussion and conclusions}

(1) I have reanalyzed the evidence that high-z supernovae support a universe with positive $\Lambda$.

(2) Both high-z supernova teams appear to have underestimated host galaxy extinction.

(3) The evidence for an $M_B - \Delta m_{15}$ relation is weaker than previously stated (only 2.6 $\sigma$)
if analysis is restricted
to supernovae observed before maximum.  The rms deviation about the mean relation is significantly
larger than previously claimed.  

(4) After consistent corrections for extinction are applied the significance of the difference in absolute
magnitude between high and low z supernovae, in an Einstein de Sitter ($\Omega_o$ = 1) universe, is
2.8-4.6 $\sigma$, depending whether (and how) the $M_B - \Delta M_{15}$ correction is applied, so such a model can
not really be rejected conclusively by the present data.  

(5) The Hubble diagram based on all high redshift estimates supports an Einstein de Sitter universe.
The HDF-N supernova favours such a universe also, contrary to the published claims of Riess et al
(2001).

(6) The community may have been too hasty in its acceptance of a positive $\Lambda$ universe, for which
no physical motivation exists, and needs to reconsider the astrophysical implications of the more
natural Einstein de Sitter, $\Omega_o$ =1, model.  For the supernova method, the need is to continue
study of low z supernovae to improve understanding of extinction and of the absolute magnitude
decay-time relation, and to consider shifting towards infrared wavelengths, as advocated by
Meikle (2000), in order to reduce the effects of extinction. 

Of course the arguments presented here do not prove that $\Lambda$ = 0.  The combination of the evidence from CMB
fluctuations for a spatially flat universe with a variety of large-scale structure arguments for
$\Omega_o$ = 0.3-0.5 may still make positive $\Lambda$ models worth pursuing.  However it would seem to 
be premature to abandon consideration of other alternatives.

\section{Acknowledgements}
I thank Peter Meikle, Adam Riess, Steve Warren, Martin Haehnelt, Bruno Leibundgut, David Branch,
Eddie Baron, Richard Ellis and an anonymous referee for helpful comments or
discussions, and Mark Phillips and Bernhard Parodi for supplying machine-readable versions of their data.

\end{document}